\documentclass{emulateapj}

\usepackage{xspace}
\usepackage{graphicx}
\usepackage{rotating}
\usepackage{natbib}
\usepackage{apjfonts}
\def \inte {\textit{INTEGRAL}}
\def \xmm {$XMM$-$Newton$}
\def \nustar {$NuSTAR$}
\def \sw {$Swift$}

\def \src {IGR\,J11215--5952}

\def \hcm {\hbox {\ifmmode $ atom cm$^{-2}\else atom cm$^{-2}$\fi}}

\def \arcsec {\hbox{$^{\prime\prime}$}}
\def \chisq {$\chi ^{2}$}

\def \ATel {Astron.\ Tel.}
\def \apj {ApJ}

\def \apjl {ApJL}
\def \apjs {ApJS}
\def \aap {A\&A}

\def \mnras {MNRAS}
\def \araa {ARA\&A}

\def \ssr {Space Science Reviews}

\newcommand{\be}{\begin{equation}}

\newcommand{\ee}{\end{equation}}

\begin{document}


\title{\xmm\ and \nustar\ simultaneous X--ray observations of IGR~J11215-5952}

\author{L. Sidoli\altaffilmark{1}, 
A. Tiengo\altaffilmark{2, 1, 3},
A. Paizis\altaffilmark{1},
V. Sguera\altaffilmark{4},
S. Lotti\altaffilmark{5} and L. Natalucci\altaffilmark{5}
}

\altaffiltext{1}{INAF, Istituto di Astrofisica Spaziale e Fisica Cosmica, Via E.\ Bassini 15,   I-20133 Milano,  Italy ; sidoli@iasf-milano.inaf.it}
\altaffiltext{2}{Scuola Universitaria Superiore IUSS Pavia, piazza della Vittoria 15, I-27100 Pavia, Italy }
\altaffiltext{3}{INFN, Sezione di Pavia, via A. Bassi 6, I-27100 Pavia, Italy}
\altaffiltext{4}{INAF, Istituto di Astrofisica Spaziale e Fisica Cosmica, Via Gobetti 101, I-40129 Bologna, Italy}
\altaffiltext{5}{INAF, Istituto di Astrofisica e Planetologia Spaziali, Via Fosso del Cavaliere 100,   I-00133 Roma,  Italy }

\begin{abstract}
We report the results of an \xmm\  and \nustar\ 
coordinated observation of the Supergiant Fast X--ray Transient (SFXT) \src,
performed on  February 14, 2016, 
during the expected peak of its brief outburst, which repeats every  $\sim$165~days.
Timing and spectral analysis were performed simultaneously in the energy band 0.4--78 keV.
A spin period of 187.0~($\pm{0.4}$)~s was measured, consistent with previous 
observations performed in 2007. 
The X--ray intensity shows a large variability (more than one order of magnitude) 
on timescales longer than the spin period, with several luminous X--ray flares  which repeat every 2-2.5 ks, 
some of which simultaneously observed by both satellites.
The broad-band (0.4-78 keV) time-averaged spectrum was well deconvolved with a double-component model (a blackbody plus a power-law with
a high energy cutoff) together with a weak iron line in emission at 6.4 keV (equivalent width, EW, of 40$\pm{10}$~eV). 
Alternatively, a partial covering model also resulted in an adequate description of the data.
The source time-averaged X--ray luminosity was 10$^{36}$~erg~s$^{-1}$ (0.1-100 keV; assuming 7 kpc). 
We discuss the results of these observations in the framework of the different models 
proposed to explain SFXTs,  supporting a quasi-spherical settling accretion regime, 
although alternative possibilities (e.g. centrifugal barrier) cannot be ruled out.

\end{abstract}



        \section{Introduction\label{intro}}

The last decade has witnessed an important advance in understanding high mass X--ray binaries (HMXBs),
thanks to the discovery of Supergiant Fast X--ray Transients (SFXTs; \citealt{Sguera2005, Negueruela2005b})
during \inte\ \citep{Winkler2003, Winkler2011} monitoring observations of 
the Galactic plane at hard X--ray energies, above 20 keV.

SFXTs are celestial objects characterized by outbursts of brief duration (with the brightest phase typically 
concentrated in one day) made of a sequence of luminous X--ray flares (normally reaching 10$^{36}$-10$^{37}$~erg~s$^{-1}$), 
each flare lasting a few thousand seconds at most. 
SFXTs are associated with early-type supergiants, making them a subclass of HMXBs.
Their X--ray properties were quickly recognized as a challenge to the accretion models:
apparently, SFXTs share similar properties (donor spectral type, orbital geometry, neutron star spin periods) 
to classical persistent HMXBs hosting supergiant stars, 
except for their extremely variable X--ray behaviour (see, e.g., \citealt{Sidoli2013review, Walter2015}).

Their bright X--ray flares are typically sporadic and rare, occurring with a 
very low duty cycle which ranges from 0.01\% to 5\% at most \citep{Paizis2014}.
The time-averaged X--ray luminosity is below 10$^{34}$~erg~s$^{-1}$ \citep{Sidoli2008:sfxtspaperI, Bozzo2015}.

Although some SFXTs with measured orbital periods show outbursts more frequently near periastron, 
the occurrence of an SFXT flare in most cases cannot be predicted. 
The exception is \src, which is the only source of the class
that undergoes  outbursts periodically \citep{SidoliPM2006}, every $\sim$165 days \citep{Sidoli2007, Romano2009}.
This periodicity is interpreted as due to the orbital period of the system.
\citet{Sidoli2007} suggested that its clocked and brief outbursts are produced by enhanced accretion when
the neutron star of the system crosses the focussed 
outflowing wind along a plane inclined with respect to the orbit near periastron passage.

Its optical counterpart, HD~306414, is a B0.5~Ia star, located at a distance $d\ga7\:$kpc \citep{Lorenzo2014}.
The high resolution optical spectroscopy suggests a high orbital eccentricity  (e$>$0.8, \citealt{Lorenzo2014}).
\src\ is also an X--ray pulsar (P$_{spin}$=187~s; \citealt{Swank2007}), a property which makes this SFXT a laboratory to unveil the accretion mechanism
driving the SFXT behaviour. 

In this paper, we report on an \xmm\ observation coordinated with \nustar\ of \src\ performed in February 2016, 
aimed at investigating the broad-band spectrum searching for cyclotron absorption lines, to directly measure the neutron star magnetic field, and disentangle
between competing models proposed to explain the X-ray behaviour of SFXTs. 
A monitoring with \sw\ allowed us to place these observations within the context of the whole outburst. 
We assume a  source distance of 7~kpc \citep{Lorenzo2014}.

 	 \section{Observations and Data Reduction}
         \label{data_redu}

\subsection{\xmm}

The \xmm\ $Observatory$ \citep{Jansen2001} carries three X--ray
telescopes, each with an European Photon Imaging Camera (EPIC; 0.4-12 keV)
at the focus: two with MOS CCDs \citep{Turner2001} and one with a pn CCD
\citep{Struder2001}. Reflection Grating Spectrometer (RGS; 0.4-2.1 keV)
arrays \citep{DenHerder2001} are placed at two of the telescopes.

The \xmm\ observation targeted on \src\ started on 2016 February 14 at 21:25 and ended the day after at 03:36, 
resulting in net exposure times of 22 ks and 15.5~ks for the MOS and the pn, respectively. 
All EPIC cameras were operated in small window mode and adopted the medium filter.

We reduced the data using version 15.0.0 of the Science Analysis Software (SAS) 
with standard procedures and the most updated calibration files. 
We used the {\em HEASoft} version 6.18 to perform the data analysis.

At first, we extracted EPIC source counts from circular regions
of 30\arcsec\ radius, adopting PATTERN selection from 0 to 4 in the pn 
and from 0 to 12 in MOS spectra.
We evaluated the possible presence of pile-up in EPIC data with {\em epatplot}, finding moderate pile-up only in MOS data.
This suggested to excise the core of the PSF in both MOS data, extracting MOS1 and MOS2 spectra
from annular regions with inner radius of 5\arcsec and an outer radius of 30\arcsec.
Background spectra were obtained from similar sized regions offset from the source positions. 
The full exposure time was exploited and no further filtering was applied to the data.

The SAS task {\em rgsproc} was used to
extract the RGS1 and RGS2 spectra, which were later combined into one single grating spectrum
using {\em rgscombine} present in the SAS.

Appropriate response matrices were generated using the SAS tasks {\em arfgen} and {\em rmfgen}.
EPIC spectra were analysed in the energy range 0.4-12, while the RGSs, which operated in spectroscopy mode \citep{DenHerder2001}, 
were analysed in the range 0.4-2.1~keV.

To ensure applicability of the \chisq\ statistics, the spectra were rebinned such that at least 20 counts per
bin were present.
Free relative normalizations between the spectra were included to account for uncertainties in
instrumental responses, always fixing at 1 the normalization of the EPIC pn spectrum.
Spectral uncertainties are given at 90\% confidence for one interesting parameter.

In the spectral fitting we used the photoelectric absorption
model {\sc TBabs} in {\sc xspec}  adopting the interstellar abundances of \citet{Wilms2000}
and photoelectric absorption cross-sections by \citet{bcmc}.

The final, time-averaged, source net count rates were the following:
5.67$\pm{0.02}$  counts~s$^{-1}$ (EPIC pn), 
1.143$\pm{0.007}$ counts~s$^{-1}$  (MOS1, PSF core excised),
1.130$\pm{0.007}$ counts~s$^{-1}$ (MOS2, PSF core excised),
0.027$\pm{0.001}$ counts~s$^{-1}$  (combined RGS1 and RGS2 1st order spectra).

The high absorption towards the source (column density, N$_{H}$, around 10$^{22}$~cm$^{-2}$)
implied significant detection in the RGSs only above 1 keV. This limited the available RGS band to the range 1-2 keV.
The low statistics leave us with an apparent featureless RGS spectrum. 
Therefore we will not discuss it further in the paper.

\subsection{\nustar}

\src\ was observed by \nustar\ as part of the joint \xmm/\nustar\ programme
approved during  \xmm\ Announcement of Opportunity AO14, with 
the main aim of  searching for cyclotron lines.

\nustar\ observations started on 2016 February 14 at 17:11 and ended the day after at 05:01,
for a net exposure time of 20.3~ks.

The Nuclear Spectroscopic Telescope Array
(\nustar, \citealt{Harrison2013}) is the first focusing hard X-ray telescope 
above 10 keV and was launched in 2012. 
It carries two identical co-aligned telescopes that focus X-ray photons onto two independent
Focal Plane Modules, called A and B (hereafter FPMA and FPMB).
Each FPM contains four (2$\times$2) solid-state cadmium zinc telluride (CdZnTe) pixel detectors
which provide a spatial resolution of 18$''$ (full width at half maximum, FWHM), a spectral resolution of 
400~eV (FWHM) at 10 keV, and a 12$'$$\times$12$'$ field-of-view (FOV).
The effective area is calibrated in the energy range 3-78~keV \citep{Madsen2015}.
The low satellite orbit (with a $\sim$90~minute period) produces data-gaps lasting about 30~minutes 
every orbit in the \nustar\ observations.

We processed \nustar\ data (obsID 30101008002) with the version 1.5.1 (2015, June 9) 
of the \nustar\ data analysis software ($NuSTAR~DAS$),
using CALDB version 20150316. 
Spectra and lightcurves were extracted with {\em nuproducts} 
on the cleaned event files. Circular extraction regions with a  60$''$ radius were used around \src\ and for the background, 
well away from the point-spread function (PSF) source wings.
\nustar\ observation was not contaminated by any stray-light produced by sources outside the FOV.
Since the background was stable and constant along the observation, no further filtering was applied.
\nustar\ source light curves with arrival times corrected to the Solar System barycenter have been extracted
with the $NuSTAR~DAS$ tool {\em nuproducts} and ``barycorr=yes''.
When needed, good-time intervals (GTIs)  were generated using {\em xselect} and then running 
{\em nuproducts} using the ``usrgtifile'' keyword, to correctly extract the temporal selected spectra.

The  source net count rates  were the following (3-78 keV):
1.783$\pm{0.009}$ counts~s$^{-1}$  (FPMA) and 
1.678$\pm{0.009}$ counts~s$^{-1}$   (FPMB).
Spectra were rebinned in order to have at least 20 counts per bin
to ensure the applicability of the $\chi ^2$ statistics.
Spectra from FPMA and FPMB were simultaneously fitted in {\sc xspec}, without merging them together.
Cross-calibration constants were used during the spectral analysis to take into account 
calibration uncertainties.

The same procedure was used when \xmm\ and \nustar\ spectra were jointly fitted together.

\subsection{\sw}

After acceptance of the joint \xmm/\nustar\ AO14 observations,
we asked for a monitoring with \sw\ around 2016, February 14.
Five snapshots (2~ks duration each), from 2016, February 11 to February 19 were performed,
with the main aim of monitoring the source light curve to put the \xmm\ and \nustar\ \src\ intensity
into context of the outburst evolution.

\begin{figure}
\begin{center}
\centerline{\includegraphics[width=5.7cm,angle=-90]{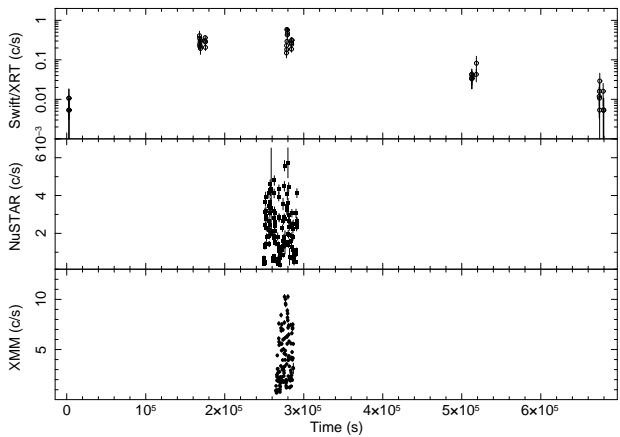}} 
\caption{\src\ light curves during the 2016 February outburst, observed by three different satellites: 
\sw/XRT (PC mode, 0.3-10 keV), 
\nustar\ (FPMA, 3-78 keV) and \xmm\ (EPIC pn, 0.4-12 keV). 
Time is in units of seconds from the first \sw/XRT observation performed on February 11.
}
\label{fig:3sat}
\end{center}
\end{figure}

The \sw/XRT source light curve (Fig.~\ref{fig:3sat}, upper panel) 
was built making use of the XRT data products generator available at the 
UK \sw\ Science Data Centre \citep{Evans2007}.

  	\section{Analysis and Results\label{result}}

\subsection{Light curves}
\label{sec:lc}

In Fig.~\ref{fig:3sat} we show the source light curve observed by \sw, $NuSTAR$ and \xmm\  during the  outburst. 
In particular, the \sw\ monitoring demonstrates that both $NuSTAR$ and \xmm\ observations caught the transient at its peak, as expected.

\begin{figure*}
\begin{center}
\centerline{\includegraphics[width=15.0cm,angle=-90]{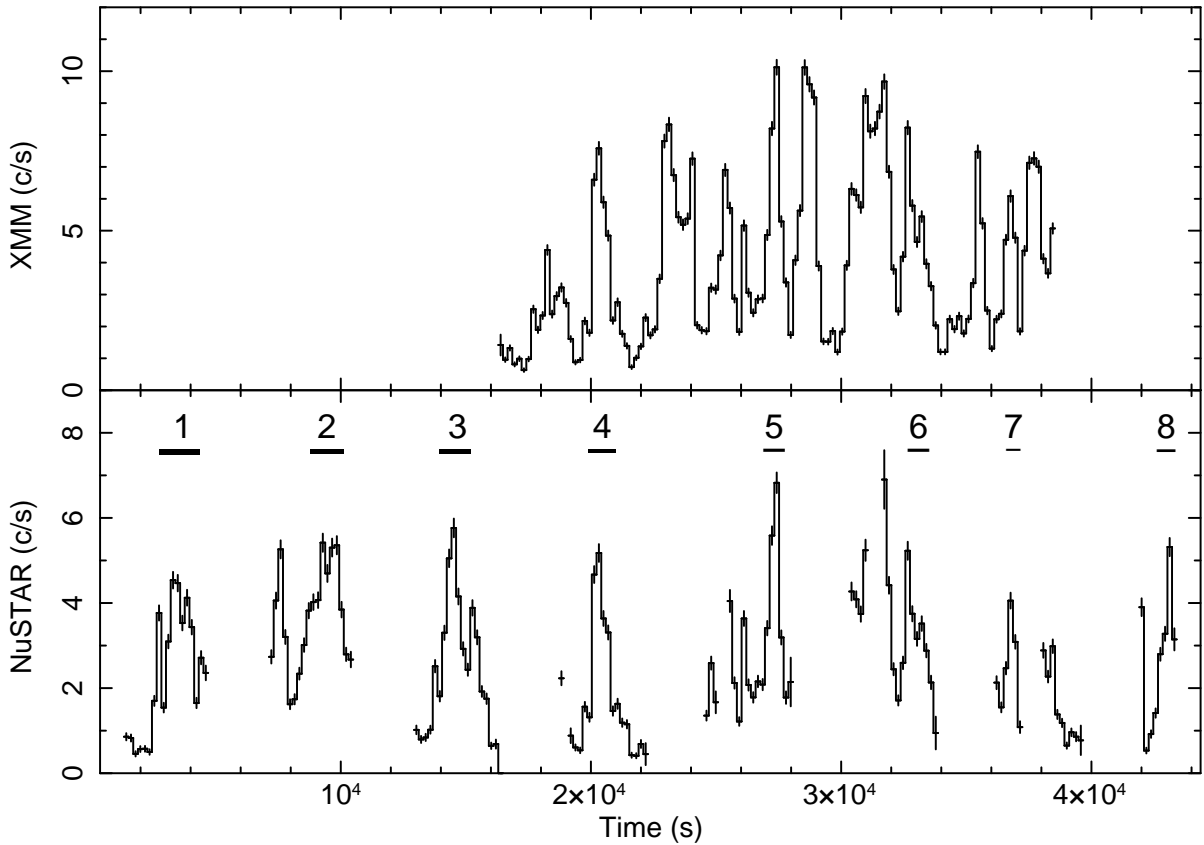}} 
\vspace{-1.5cm}
\caption{\src\ light curve during the February 2016 outburst observed by EPIC pn ({\em upper panel}; 
0.4--12 keV) and \nustar\ FPMA ({\em lower panel}; 3-78 keV). The bin time is 187~s which is its pulsation period. 
Numbers indicate the sequence of flares in \nustar\ observation (same numbers used in the temporal-selected spectroscopy).
The small horizontal lines under the numbers approximately indicate the exposure times of each flare spectrum.
}
\label{fig:pn_nu_lc}
\end{center}
\end{figure*}

In Fig.~\ref{fig:pn_nu_lc} we show only the light curves as observed by \xmm\ (EPIC pn, 0.4-12 keV)
and \nustar\ (FPMA, for clarity, 3-78 keV)  binned on the known spin (187~s, see below), to avoid variability 
due to the neutron star rotational period.
The intensity variability is large, in excess of one order of magnitude.
The uninterrupted \xmm\ light curve clearly shows that each satellite gap in the NuSTAR observation 
is actually filled by bright flares,  repeating on a timescale of 2-2.5~ks.

We performed  timing analysis on EPIC (all three cameras, 2-12 keV) and \nustar\ data (FPMA and FPMB, 3-78 keV)
after correcting arrival times to the Solar System barycenter, applying  epoch folding techniques \citep{Leahy1983}
to measure the known neutron star spin period. 
We determined a periodicity of 187.0$\pm{0.4}$~s,
consistent with what observed in 2007
with \xmm\ \citep{Sidoli2007} and with $RXTE$/PCA \citep{Swank2007}.

\begin{figure*}
\begin{center}
\begin{tabular}{cc}
\hspace{0.0cm}
\includegraphics[width=8.5cm,angle=0]{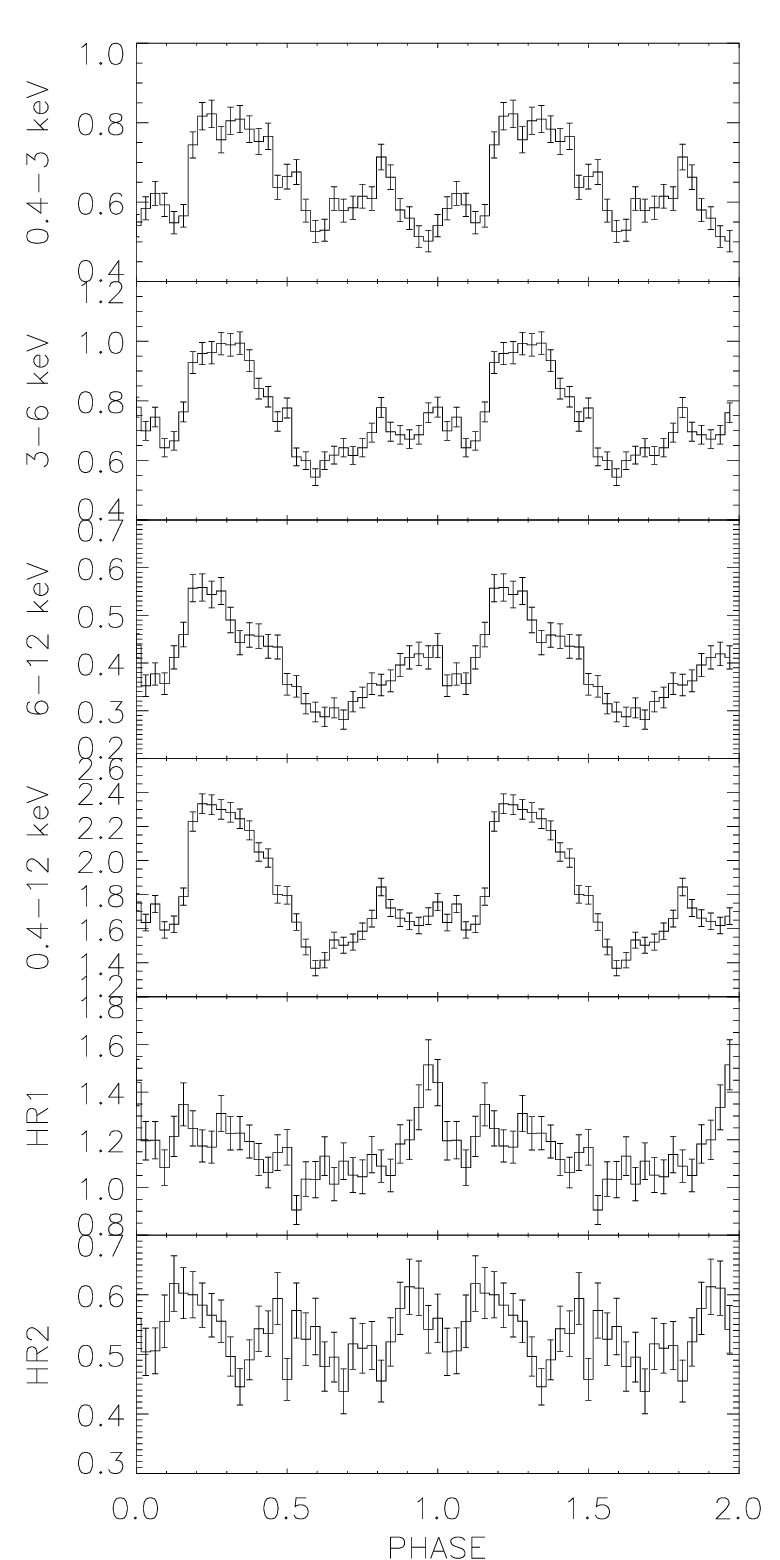}
\includegraphics[width=8.5cm,angle=0]{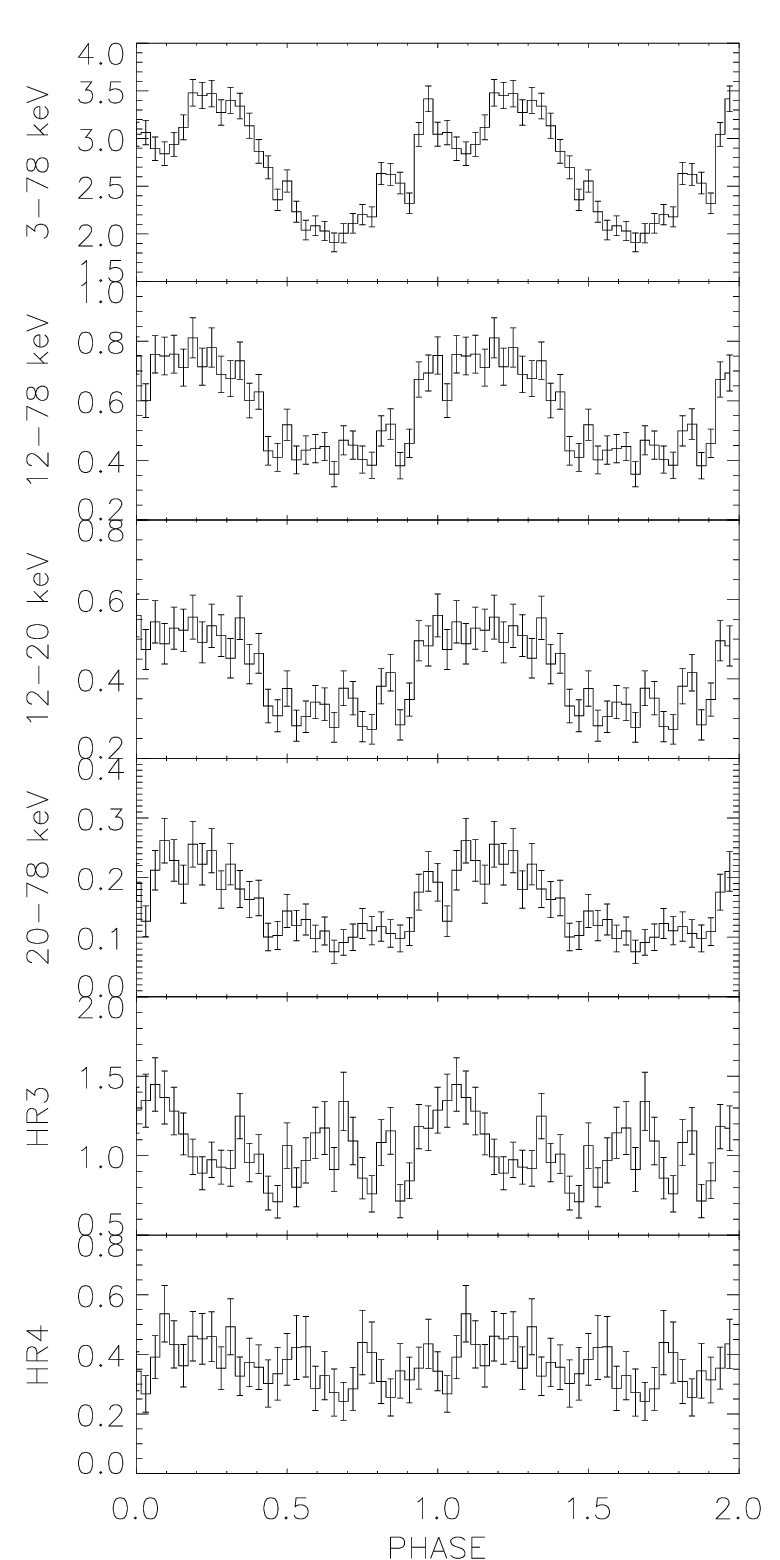}
\end{tabular}
\end{center}
\caption{\src\ pulse profiles (period P=187~s; MJD~57432 is the epoch of $\phi$=0) in different energy ranges are shown, as observed simultaneously 
by \xmm\ (EPIC pn, {\em left panels}) and by NuSTAR (FPMA for clarity, {\em right panels}).
Four hardness ratios are displayed, calculated in the following energy ranges: 
HR1=(3-6 keV)/(0.4-3 keV); 
HR2=(6-12 keV)/(3-6 keV);
HR3=(12-20 keV)/(6-12 keV);
HR4=(20-78 keV)/(12-20 keV).
}
\label{fig:spinprofiles}
\end{figure*}

In Fig.~\ref{fig:spinprofiles} we show the pulse profiles extracted from the strictly simultaneous \xmm\ and \nustar\ data,
in different energy ranges, together with four hardness ratios. The pulse profile is energy dependent,
being dominated by an asymmetric main peak below 12 keV,
which becomes symmetric at harder energies. 
The hardness ratios display different behavior along the pulse phase, as
show in the lower panels in Fig.\ref{fig:spinprofiles}.

The pulsed fraction, PF, calculated as 
\mbox{$PF\equiv(F_{\rm{max}}-F_{\rm{min}})/(F_{\rm{max}}+F_{\rm{min}})$},
where F$_{\rm max}$ and F$_{\rm min}$ are the count rates at the maximum and at the minimum of the pulse profile, respectively, 
increases with energy, from $24\pm{3}$~\% (0.4-3 keV) to $60\pm{14}$~\% (20-78 keV). 
Over the whole \xmm\ energy band PF=$26\pm{2}$~\% (0.4-12 keV), 
while PF=$29\pm{3}$~\% over the whole \nustar\ band (3-78 keV).

\subsection{Spectra}
\label{sec:spec}

\subsubsection{Time-averaged \xmm\ and \nustar\ spectrum}
\label{sec:avspec}

We extracted \nustar\ spectra overlapping in time with the whole \xmm\ observation,
to investigate the broad-band (0.4-78 keV) time-averaged \src\ emission.
This translated into a net exposure time of $\sim$9 ks for each \nustar\ FPM. 
%
\begin{figure}
\begin{center}
\centerline{\includegraphics[width=6.0cm,angle=-90]{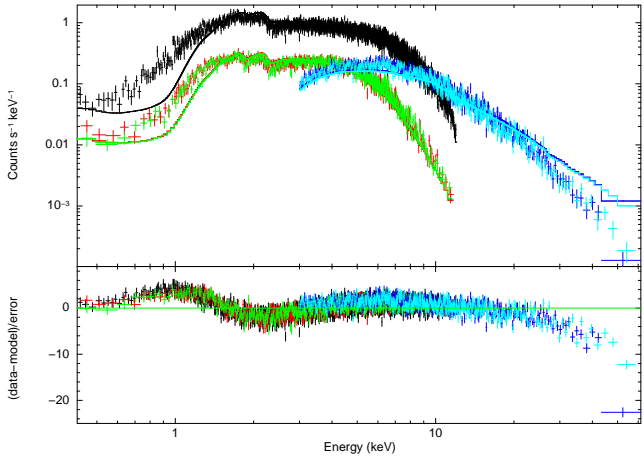}} 
\vspace{0.0cm}
\caption{
Time-average broad-band spectrum, using EPIC and \nustar\ data. 
Count spectra are shown together with the residuals
(in units of standard deviation) with respect to a simple absorbed power law, 
to clearly show the structures in the residuals.
EPIC spectra are marked in black (pn), red (MOS1) and green (MOS2) colors, 
\nustar\ spectra  in blue (FPMA) and in pale blue (FPMB).
}
\label{fig:av_spec_pow}
\end{center}
\end{figure}

\begin{figure*}
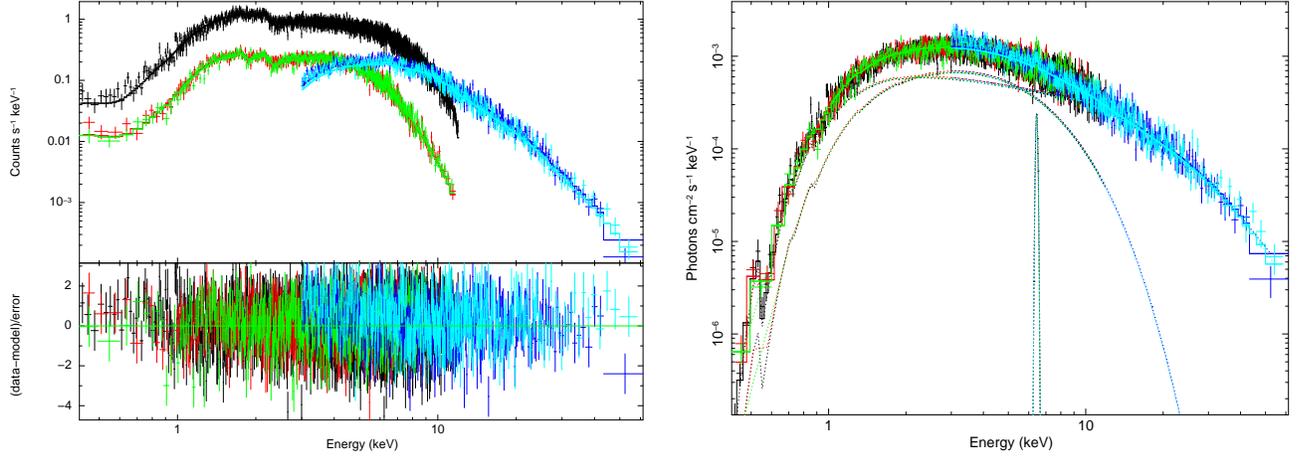

\begin{center}
\begin{tabular}{lc}
\includegraphics[width=6.cm,angle=-90]{fig05a.ps} 
\includegraphics[width=6.cm,angle=-90]{fig05b.ps}
\end{tabular}
\end{center}
\caption{Time-average broad-band spectrum, using EPIC and \nustar\ data 
(only events overlapping in time with the \xmm\ observation, as discussed in Sect.~\ref{sec:avspec}). 
Count spectra are shown on the left together with the residuals
(in units of standard deviation) with respect to the best-fit model reported in Table~\ref{tab:av_spec}. Photon spectra are displayed on the right.
EPIC spectra are marked in black (pn), red (MOS1) and green (MOS2) colors, \nustar\ spectra  in blue (FPMA) and in pale blue (FPMB).
}
\label{fig:av_spec}
\end{figure*}

The joint fit of the five spectra (EPIC pn, MOS1, MOS2, FPMA and FPMB) 
with a single absorbed power-law model provided unacceptable results
($\chi^{2}_{\nu}$=2.89 for 3292 degrees of freedom, dof; Fig.~\ref{fig:av_spec_pow}), because of an evident
rollover at hard energies, together with a soft excess (below 1.5 keV) and positive residuals at $\sim$6.4 keV, 
consistent with a K$_{\alpha}$ emission line from neutral iron. 
We tried different double-component continuum models. The
best-fit was obtained adopting a 
blackbody (with a temperature, kT$_{\rm BB}$, of $\sim$1.6~keV, and a
radius, ${\rm R_{BB}}$, of a few hundred meters, calculated assuming a distance of 7~kpc),
together with a hard power-law (photon index, $\Gamma$, of 0.6) modified by a cut-off at high energy (hereafter PLCUT), 
defined as M(E)=${ (exp[(E_{cut}-E)/E_{fold}] )}$ when E$ \ge E_{cut}\:$, while M(E)=1 at E$ \le E_{cut}\:$.
An additive Gaussian line in emission around 6.4~keV was also needed.
The results are reported in Table~\ref{tab:av_spec} (Model~1) and shown in Fig.~\ref{fig:av_spec}. 
The presence of a blackbody component was already known from previous \xmm\ observations of IGR~J11215-5952 
performed in 2007 \citep{Sidoli2007}  and it usually accounts for soft excesses detected in several
accreting pulsars \citep{LaPalombaraMereghetti06}.
We also tried an alternative decomposition of the broad-band spectrum,
replacing the blackbody component with a partial covering fraction absorption ({\sc pcfabs} model in {\sc xspec}).
This model consists of 
an additional intrinsic absorption (N$_{\rm H pcfabs}$) with a covering fraction f$_{\rm  pcfabs}$ (${\rm 0 < f_{\rm  pcfabs} < 1}$),
so that the continuum is:
\[
{\rm Intensity = e^{-\sigma N_H} \;[fe^{-\sigma N_{H pcfabs}} \rm  +\; (1 -\; f)] \; (I_{PLCUT}) ,}
\]
A partially covered PLCUT model (hereafter Model 2) provided an equivalently good description of the data. 

Note that consistent spectral results were obtained 
from strictly simulataneous \xmm\ and \nustar\ time-averaged spectra, but we considered here 
the whole \xmm\ exposure time to better constrain the parameters of the iron emission line.

\begin{table}
        \centering
\caption[]{Broad-band time-averaged spectrum (0.4--78 keV).
}
\begin{tabular}{lcc}
 \hline
\noalign {\smallskip}
Parameter                                  &     Model~1                              &       Model~2       \\
\hline
\noalign {\smallskip}
N$_{\rm H}$      ($10^{22}$~cm$^{-2}$)     &  $1.05\pm{0.06} $                        &     $1.27 ^{+0.07} _{-0.09}$        \\
 N$_{\rm H pcfabs}$ ($10^{22}$~cm$^{-2}$)  &       none                 &     $4.5 ^{+0.6} _{-0.9}$          \\
 f$_{\rm  pcfabs}$                               &        none                &        $0.51\pm{0.04} $         \\
kT$_{\rm BB}$ (keV)                        &  $1.61 ^{+0.06} _{-0.07}$                &       none     \\
${\rm R_{BB}}$ (m)                         &    $460\pm{30}$                          &       none     \\
$\Gamma$$^a$                                  &  $0.56 ^{+0.08} _{-0.11}$              &       $0.97 ^{+0.11} _{-0.08}$              \\
E$_{cut}$ (keV)                              &  $9.7 ^{+0.9} _{-1.0}$                  &       $5.5 ^{+0.8} _{-0.4}$                  \\
E$_{fold}$ (keV)                              &  $14.3\pm{1.4}$                        &        $15.3^{+1.8} _{-1.4}  $      \\
E$_{line}$  (keV)                          &  $6.44 ^{+0.03} _{-0.03}$                  &     $6.45 ^{+0.03} _{-0.03}$           \\
$\sigma$$_{line}$ (keV)                    &  $<0.09$                                   &       $<0.09$      \\
Flux$_{line}$ (ph~cm$^{-2}$~s$^{-1}$)      &  ($3.2\pm{0.8}$)$\times10^{-5}$            &     ($3.2 ^{+0.9} _{-0.4}$)$\times10^{-5}$      \\
EW$_{line}$$^b$  (eV)                      &  $40\pm{10}$                               &      $41 ^{+12} _{-5}$      \\
Flux$^c$      (erg~cm$^{-2}$~s$^{-1}$)         &    (1.7$\pm{0.1}$) $\times$10$^{-10}$    &    (1.8$\pm{0.1}$) $\times$10$^{-10}$        \\
Luminosity$^c$  (erg~s$^{-1}$)                 &    (1.0$\pm{0.1}$)   $\times$10$^{36}$   &       (1.1$\pm{0.1}$)   $\times$10$^{36}$     \\
$\chi^{2}_{\nu}$/dof                       &    1.042/3285                                &        1.057/3285     \\
\hline
\label{tab:av_spec}
\end{tabular}
\tablecomments{
$^a$ Power-law photon index. \\
$^b$ Equivalent width. \\ 
$^c$ Fluxes (corrected for the absorption) and luminosities are in the energy range 0.1--100~keV. The flux reported for  Model~2 is corrected for both 
absorbing components.}
\end{table}

\subsubsection{Temporal-selected spectroscopy}
\label{sec:nu_revsel}

\begin{figure}
\begin{center}
\begin{tabular}{cc}
\includegraphics[width=5.5cm,angle=-90]{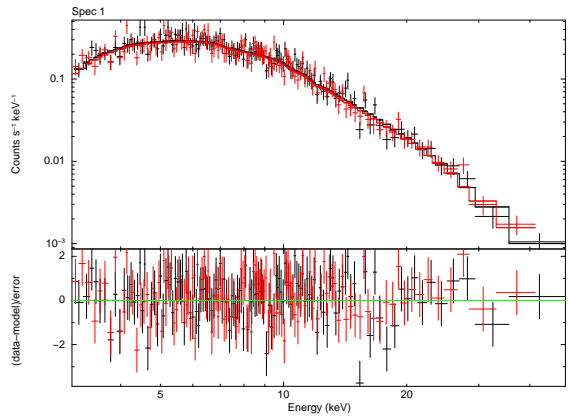} 
\end{tabular}
\end{center}
\caption{\nustar\ temporal-selected spectroscopy (FPMA in black, FPMB in red) of the  peak of flare n.~1.
Counts spectrum is shown (together with the residuals), fitted with an absorbed PLCUT model (Table~\ref{tab:revsel}).
}
\label{fig:revsel_plcut}
\end{figure}

Cyclotron resonant scattering features (CRSFs) in X--ray pulsars with standard 
neutron star magnetic fields (a few $10^{12}$~Gauss) can be detected in the X--ray spectrum at energies E$>$10~keV. 
Thus, in principle, if they are strong enough also in \src\, they should be found in the \nustar\ spectra alone.
Therefore, we started concentrating on \nustar\ data only, extracting spectra from the peaks of the flares, 
as shown in Fig.~\ref{fig:pn_nu_lc} (numbers in the lower panel).

When fitting \nustar\ data only, the soft blackbody component that was evident 
in the broad-band time-averaged spectrum (Sect.~\ref{sec:avspec})
could not be constrained, as well as the absorption column density and the iron emission line (this latter is evident only in the time-averaged spectrum). 
Since we were interested in searching for absorption features at hard energies (E$>$10 keV), 
we preferred not to include the blackbody component and the iron line in the spectral model. 
The absorbing column density was a free parameter in the fit. 
In this manner, it naturally converged to very high (and unrealistic) values, 
with respect to the time-averaged spectroscopy of \xmm\ and \nustar\ simultaneous data. 
As a first attempt, we simply adopted this absorbed PLCUT model as a pure phenomenological and statistically acceptable description of the continuum emission.
The results from fitting  the eight time-resolved 3-78 keV spectra in this way are reported in Table~\ref{tab:revsel}.

\begin{table*}
        \centering
\caption[]{Time-resolved spectroscopy (3--78 keV),  fitting only \nustar\ data  of the flare peaks. 
}
\begin{tabular}{lcccccccc}
 \hline
Parameter                                     &  Spec.   1               &  2                  &  3                       &  4                           &         5                 &      6                      &        7                      &     8        \\
\hline\noalign {\smallskip}
N$_{\rm H}$      ($10^{22}$~cm$^{-2}$)        &  $  7^{+5} _{-4}$        &  $  8 \pm{3}$       &    $6 \pm{3}$            &   $5 ^{+6} _{-5}$            &     $9\pm{4}$             &   $9 \pm{4}$                &   $5 ^{+6} _{-5}$             &  $3 ^{+6} _{-3}$        \\
$\Gamma$                                  &  $1.4 ^{+0.5} _{-0.3}$  &  $1.6 ^{+0.2} _{-0.2}$ & $1.04 ^{+0.22} _{-0.22}$ &  $1.27 ^{+0.37} _{-0.30}$ &  $1.69 ^{+0.15} _{-0.61}$ &  $1.64 ^{+0.11} _{-0.10} $  &   $1.6 ^{+0.4} _{-0.7}$       &   $1.28 ^{+0.19} _{-0.60}$   \\
E$_{cut}$ (keV)                               &  $6 \pm{6}$              &  $7 ^{+3} _{-2}$    &  $3 ^{+4} _{-3}$         &  $7 ^{+5} _{-7}$             &  $15 ^{+8} _{-15}$        & $22 ^{+13} _{-3}$           &  $9 ^{+14} _{-9}$             &    $8 ^{+5} _{-8}$      \\
E$_{fold}$ (keV)                              &  $27 ^{+35} _{-8}$       &  $33 ^{+20} _{-13}$ &  $15 ^{+5} _{-3}$        &  $25 ^{+22} _{-8}$           &  $40 ^{+50} _{-20}$       & $19 ^{+28} _{-19}$          &  $30^{+40} _{-12} $           &   $20 ^{+9} _{-9}$     \\
F (10$^{-10}$ erg~cm$^{-2}$~s$^{-1}$)$^a$      &    3.0$\pm{0.3}$        &  3.6$\pm{0.4}$      &   2.5$\pm{0.2}$          &  3.2$\pm{0.3}$               &  4.6$\pm{0.3}$            & 3.1$\pm{0.2}$                &    2.3$\pm{0.7}$             &    1.7$\pm{0.2}$       \\
L  (10$^{36}$ erg~s$^{-1}$) $^a$               &    1.8$\pm{0.2}$        &  2.1$\pm{0.2}$      &   1.5$\pm{0.1}$          &  1.9$\pm{0.2}$               &  2.7$\pm{0.2}$            & 1.8$\pm{0.1}$                &    1.3$\pm{0.4}$             &    1.1$\pm{0.1}$       \\
$\chi^{2}_{\nu}$/dof                           &    1.046/239            &  0.949/347          &    0.998/282             &  1.121/155                   &  0.900/164                &  0.887/160                   &     0.602/89                 &   0.839/94             \\
\hline
\label{tab:revsel}
\end{tabular}
\tablecomments{
$^a$ Fluxes (corrected for the absorption) and luminosities are in the energy range 1--100~keV.
}
\end{table*}

\begin{table*}
        \centering
\caption[]{Same as Table~\ref{tab:revsel} (\nustar\ only), considering only flare peaks 1,2,3,4 and 6, where the inclusion of an absorption feature resulted in a line significance larger than 2~$\sigma$.
}
\begin{tabular}{lccccc}
 \hline
Parameter                                  &  Spec.   1                   &  2                        &  3                           &  4                       &  6               \\
\hline\noalign {\smallskip}
N$_{\rm H}$      ($10^{22}$~cm$^{-2}$)     &  $  10 \pm{3}$       &  $7 ^{+4} _{-7}$          &    $6 \pm{4}$                &   $4 ^{+6} _{-4}$                &   $6 ^{+4} _{-4}$      \\
$\Gamma$                                   &  $1.6 ^{+0.2} _{-0.2}$  &  $1.40 ^{+0.19} _{-1.39}$ & $1.4 ^{+0.3} _{-0.4}$        &  $1.15 ^{+0.31} _{-0.32}$     &   $1.49 ^{+0.14} _{-0.14}$   \\
E$_{cut}$ (keV)                            &  $19 ^{+4} _{-9}$       &  $6.3 ^{+2.2} _{-6.3}$    &  $7 ^{+4} _{-7}$             &  $6.4 ^{+1.8} _{-6.4}$        &   $20 ^{+3} _{-3}$           \\
E$_{fold}$ (keV)                           &  $21 ^{+21} _{-8}$      &  $27 ^{+16} _{-13}$       &  $31 ^{+95} _{-15}$          &  $23 ^{+12} _{-7}$            &   $15 ^{+9} _{-5}$           \\
Depth$_{cycl}$$^a$                             &  $0.44 ^{+0.18} _{-0.17}$  & $0.16 ^{+0.16} _{-0.11}$ & $0.74 ^{+2.9} _{-0.46}$    &  $0.38 ^{+0.31} _{-0.26}$     &   $0.46 ^{+0.27} _{-0.25}$   \\
E$_{cycl}$$^a$  (keV)                           &  $16.7 ^{+1.0} _{-0.9}$  &  $15 ^{+3} _{-5}$          &  $32.5 ^{+4.2} _{-3.4}$    &  $18.0 \pm{1.0}$              &    $17.2 \pm{0.7}$          \\
Width$_{cycl}$$^a$  (keV)                       &  $2.6 ^{+3.2} _{-1.5}$   &  $1.6 ^{+14} _{-1.6}$      &  $4.9 ^{+11.3} _{-4.9}$    &  $1.0 ^{+2.4} _{-1.0}$        &      $1.0 ^{+1.2} _{-1.0}$   \\
Line Signif.$^b$                               &    4.4~$\sigma$           &    2.3~$\sigma$         &       2.7~$\sigma$          &  2.5~$\sigma$                  &  3.25~$\sigma$      \\
$\chi^{2}_{\nu}$/dof                       &    1.012/236              &  0.942/344               &    0.984/279                &  1.103/152                    &  0.837/157              \\
\hline
\label{tab:revsel_cyclabs}
\end{tabular}
\tablecomments{
$^a$ {\sc cyclabs} model in {\sc xspec}. \\
$^b$ This line significance was calculated by increasing the $\Delta\chi^2$ value until the confidence region boundary of the depth (Depth$_{cycl}$) of the cyclotron line crosses 0.
}
\end{table*}

In some of these spectra, negative residuals appeared at energies between 15 and 40 keV.
To account for these absorption features, we modified the PLCUT 
continuum multiplying it with  a cyclotron absorption 
model ({\sc cyclabs} in  {\sc xspec} package).
At first, we calculated the line significance by increasing the $\Delta\chi^2$ value until the 
confidence region boundary of the depth (Depth$_{cycl}$ in Table~\ref{tab:revsel_cyclabs}) of the cyclotron line crosses 0.
The significance of these lines was  greater than 2$~\sigma$ in five (of eight) spectra (flares number 1, 2, 3, 4 and 6),
so we have reported absorption line parameters only for these  spectra in Table~\ref{tab:revsel_cyclabs} for clarity.
The absorption lines were measured with a significance ranging from $\sigma$=2.3 to $\sigma$=4.4 (with {\sc cyclabs} model),
with the best value  obtained 
during the first flare observed by \nustar.
The centroid energy was different: in spectra n.~1, 4 and 6 the feature was in a narrow range around 17-18 keV,
while during spectrum n.~3 it was at $\sim$33~keV.
In spectrum n.~3 the line at $\sim$17~keV was undetected: we could place an upper limit  (Depth$_{cycl}$$\la0.18\:$)
to the depth of the {\sc cyclabs} line 
with the centroid energy fixed at 17~keV (and line width of 2 keV).

In order to test the significance of the absorption line, 
since the F-test could bring unreliable results \citep{Protassov2002}, 
we exploited Monte Carlo simulations to confirm the presence of the cyclotron line in spectrum n.1 (our best case,  Fig.~\ref{fig:revsel_plcut}), 
following the approach described by several authors \citep{Bhalerao2015, Lotti2016, Barriere2015}. 
We used the \emph{simftest} routine in {\sc xspec} with 100000 simulations\footnote{The \emph{simftest} is a script that tests for the presence of an additional component 
simulating N fake spectra based on the model without such component. It then fits the spectra with both the models with and without the additional component registering 
the $\Delta\chi^{2}$ and comparing it with the $\Delta\chi^{2}$ obtained with the real data to obtain the probability of a false detection.}, 
fixing the $N_{\rm H}$ to the values reported in Table~\ref{tab:revsel_cyclabs}  
to speed up the process, and found that the probability of such line to be required by the data is 
$P=99.15\%$ for spectrum n. 1, corresponding to $2.63$~$\sigma$ (normalized to 8 trials).
The significance of the line did not improve if a blackbody component was included in the fitting model to flare n.1 spectrum, 
with a temperature, radius and absorption fixed to the broad-band time-averaged values.

\begin{table*}
        \centering
\caption[]{Broad-band (0.4-78 keV) spectroscopy of the flares simultaneously observed by \xmm\ and \nustar. 
}
\begin{tabular}{lcc|c|cc|c}
 \hline
 Model 1$^a$ Parameters                                  &    flare   4                                 &            4                  &      5                            &        6                      &             6                     &      7     \\
\hline\noalign {\smallskip}
N$_{\rm H}$      ($10^{22}$~cm$^{-2}$)     &  $1.08 ^{+0.28} _{-0.29}$               &  $0.94 ^{+0.31} _{-0.30}$     &    $0.89 ^{+0.25} _{-0.24}$        &  $1.15 ^{+0.21} _{-0.20}$    &      $1.00 ^{+0.24} _{-0.24}$     &     $1.3 ^{+0.1} _{-0.1}$      \\
kT$_{\rm BB}$ (keV)                        &  $1.68 ^{+0.43} _{-0.22}$               & $1.47 ^{+0.27} _{-0.18}$      &   $1.51 ^{+0.39} _{-0.16}$         &    $1.88 ^{+0.13} _{-0.10}$  &      $1.71 ^{+0.15} _{-0.12}$      &     $1.94 ^{+0.07} _{-0.07}$      \\
${\rm R_{BB}}$ (m)                         &    $480 ^{+180} _{-120} $              &     $600 ^{+250} _{-160}$      &    $630 ^{+200} _{-120}$           &   $450 ^{+60} _{-50}$       &         $520 ^{+90} _{-80}$          &       $380 ^{+10} _{-10}$       \\
$\Gamma$                                   &  $0.54 ^{+0.23} _{-0.36}$               &  $0.29 ^{+0.31} _{-0.94}$     & $0.43 ^{+0.25} _{-0.40}$           &   $0.85 ^{+0.13} _{-0.14}$  &        $0.64 ^{+0.19} _{-0.29}$     &     $1.11 ^{+0.05} _{-0.05}$   \\
E$_{cut}$ (keV)                           &  $6 \pm{6}$                              &  $6.4 ^{+1.2} _{-1.6}$        &  $7 ^{+11} _{-7}$                  &    $21 ^{+2.5} _{-2.2}$     &       $18.6 ^{+2.3} _{-3.2}$        &        $29 ^{+3} _{-2}$   \\
E$_{fold}$ (keV)                          &  $15 ^{+6} _{-4}$                        &  $12 ^{+4} _{-4}$             &  $13 ^{+5} _{-3}$                  &    $12 ^{+7} _{-4}$         &       $9.9 ^{+3.5} _{-2.5}$          &        $<25$      \\
Depth$_{cycl}$                             &  none                                   & $0.45 ^{+0.30} _{-0.28}$      &            none                    &      none                   &        $0.64 ^{+0.32} _{-0.33}$      &        none             \\ 
E$_{cycl}$ (keV)                           &  none                                   &  $18.1 ^{+1.0} _{-1.8}$       &            none                    &        none                 &        $17.4 ^{+0.64} _{-0.67}$      &         none       \\ 
Width$_{cycl}$ (keV)                       &  none                                   &  $1 ^{+4} _{-1}$              &             none                   &          none               &          $1 ^{+1.4} _{-1}$          &       none       \\ 
Line Signif.$^b$                           &    none                                &    2.8~$\sigma$                &       none                         &       none                   &          3.35~$\sigma$              &          none         \\ 
F (10$^{-10}$ erg~cm$^{-2}$~s$^{-1}$)$^c$   &    2.8$\pm{0.4}$                      &  2.8$\pm{0.4}$                  &   3.1$\pm{0.4}$                  &  2.4$\pm{0.2}$                &  2.4$\pm{0.2}$                      & 1.6$\pm{0.1}$                 \\
L  (10$^{36}$ erg~s$^{-1}$) $^c$            &    1.6$\pm{0.2}$                      &  1.6$\pm{0.4}$                  &   1.8$\pm{0.2}$                  &  1.4$\pm{0.1}$                &  1.4$\pm{0.1}$                       & 0.94$\pm{0.06}$                \\
$\chi^{2}_{\nu}$/dof                       &    0.989/457                            &  0.978/454                     &    1.018/503                     &  0.869/492                     &   0.851/489                         &    0.929/271                \\
\hline
Model 2$^a$ Parameters                                 &      flare  4                                 &            4                  &      5                            &        6                      &             6                     &      7     \\
\hline\noalign {\smallskip}
N$_{\rm H}$      ($10^{22}$~cm$^{-2}$)     &  $1.47 ^{+0.28} _{-0.38}$               &  $1.37 ^{+0.30} _{-0.42}$     &    $1.30 ^{+0.28} _{-0.43}$       &  $1.37 ^{+0.24} _{-0.26}$    &      $1.6 ^{+0.2} _{-0.2}$         &     $1.52 ^{+0.38} _{-0.79}$      \\
 N$_{\rm H pcfabs}$ ($10^{22}$~cm$^{-2}$)   &  $8 \pm{4}$                            & $7 ^{+4} _{-3}$               &   $7 \pm{4}$                      &  $8 ^{+6} _{-4}$             &      $10.6 ^{+2.8} _{-2.3}$        &     $10 ^{+14} _{-9}$ \\
 f$_{\rm  pcfabs}$                          &   $0.55 ^{+0.09} _{-0.14} $            & $0.51 ^{+0.12} _{-0.13}$      &   $0.55 ^{+0.08} _{-0.13}$        &  $0.49 ^{+0.18} _{-0.15}$    &      $0.66 ^{+0.04} _{-0.05}$      &     $0.4 ^{+0.4} _{-0.3}$         \\ 
$\Gamma$                                   &  $1.16 ^{+0.20} _{-0.24}$               &  $1.05 ^{+0.24} _{-0.21}$     & $1.16 ^{+0.19} _{-0.27}$           &   $1.04 ^{+0.40} _{-0.24}$  &        $1.45 ^{+0.10} _{-0.10}$    &     $1.1 ^{+0.6} _{-0.4}$   \\
E$_{cut}$ (keV)                            &  $6.7 ^{+1.2} _{-1.3}$                 &  $6.2 ^{+1.4} _{-1.1}$        &  $6.7 ^{+1.0} _{-1.5}$              &    $5 ^{+4} _{-5}$          &       $20 ^{+2} _{-3}$             &        $5 ^{+5} _{-5}$   \\
E$_{fold}$ (keV)                           &  $20 \pm{5}$                           &  $20 ^{+6} _{-4}$             &  $20 ^{+7} _{-6}$                   &    $18 ^{+16} _{-5}$         &       $15 ^{+8} _{-5}$            &       $15 ^{+20} _{-5}$    \\
Depth$_{cycl}$                             &  none                                   & $0.43 ^{+0.28} _{-0.29}$      &            none                    &      none                   &        $0.52 ^{+0.26} _{-0.26}$    &        none             \\ 
E$_{cycl}$ (keV)                           &  none                                   &  $18.0 ^{+1.0} _{-1.8}$       &            none                    &        none                 &        $17.2 ^{+0.6} _{-0.7}$      &         none       \\ 
Width$_{cycl}$ (keV)                       &  none                                   &  $1 ^{+5} _{-1}$              &             none                   &          none               &          $1 ^{+1.8} _{-1}$         &       none       \\ 
Line Signif.$^b$                            &    none                                &    2.8~$\sigma$               &       none                        &       none                   &          3.87~$\sigma$              &          none         \\ 
F (10$^{-10}$ erg~cm$^{-2}$~s$^{-1}$)$^c$   &    3.17$\pm{0.11}$                      &  3.15$\pm{0.1}$              &   3.6$\pm{0.1}$                  &  2.8$\pm{0.1}$                &  2.8$\pm{0.1}$                      & 2.0$\pm{0.1}$                 \\
L  (10$^{36}$ erg~s$^{-1}$) $^c$            &    1.86$\pm{0.07}$                      &  1.85$\pm{0.06}$             &   2.1$\pm{0.06}$                  &  1.6$\pm{0.1}$                &  1.6$\pm{0.1}$                     & 1.2$\pm{0.1}$                \\
$\chi^{2}_{\nu}$/dof                       &    0.998/457                            &  0.987/454                    &    1.029/503                     &  0.871/492                     &   0.863/489                        &    0.943/271                \\
\hline
\label{tab:simflares}
\end{tabular}
\tablecomments{
$^a$ Model~1 is a double-component continuum  made of a blackbody plus a PLCUT model. Model~2 is a PLCUT model modified by a partial covering fraction absorption {\sc pcfabs}. \\
$^b$ This line significance was calculated by increasing the $\Delta\chi^2$ value until the confidence region boundary of the depth (Depth$_{cycl}$) of the cyclotron line crosses 0. \\
$^c$ Fluxes (corrected for the absorption) and luminosities are in the energy range 0.1-100 keV. 
}
\end{table*}

\begin{figure*}
\begin{center}
\begin{tabular}{cc}
\hspace{0.7cm}
\includegraphics[width=15.5cm,angle=0]{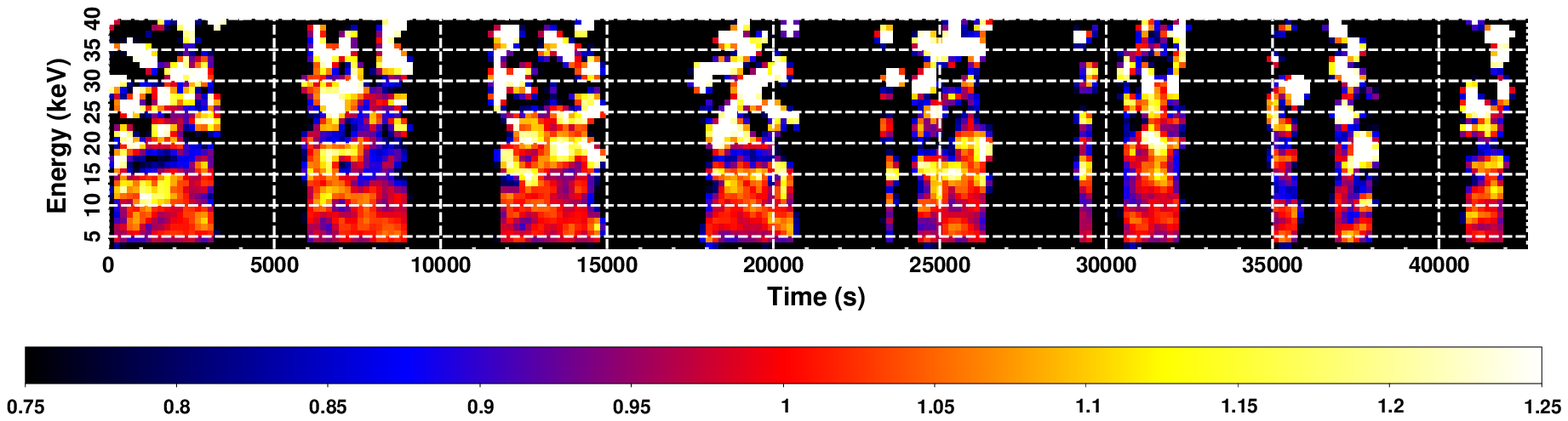}
\end{tabular}
\end{center}
\caption{Time-energy image (normalized to the time-averaged \nustar\ spectrum and light curve), using \nustar\ events in the 3-40 keV energy range. 
The color scale marks in blue/black a deficit of counts (i.e. an absorption feature) with respect to 
the time-averaged \nustar\ spectrum (whose level is at 1). The image at the highest energies is more noisy due to much lower number of counts per pixel.
}
\label{fig:entimelc}
\end{figure*}

To obtain a global view of the source behavior as observed by \nustar,
we combined together the two FPM source event files and built an ``Energy versus Time'' plot, that was then 
normalized to the time-averaged \nustar\ spectrum
and the energy-integrated light curve. Each pixel of this image represents the count rate in a 187~s time interval 
and 1~keV energy bin, divided by the time averaged count rate in the same energy bin during the whole observation 
and by the ratio of the 3-40 keV rate in the same time inteval and in the full observation.
The result is shown in Fig.~\ref{fig:entimelc}:
the different colors in the image mark the residuals of the source counts at different energies
with respect to the time-averaged spectrum and broad band light curve (color level at 1), 
along the \nustar\ observation (the gaps due to the eight satellite revolutions are clearly visible).
We note that this is the only time where we combined together the two \nustar\ FPMs (here not barycentered) events, to get better statistics.
From the visual inspection of Fig.~\ref{fig:entimelc}, some dark horizontal lines 
(deficit of counts with respect to the time-averaged \nustar\ spectrum) 
 appear in the first four satellite orbits, between 15 and 35~keV, 
confirming  the hint of absorption features  suggested by the spectral analysis.
We note that, although this plot is not able to assess the significance
of any negative residual, nevertheless it shows that we did not miss other eventual spectral
features with our particular time (and intensity) resolved spectral extraction.

A further spectral investigation was performed considering flare peaks simultaneously observed by \xmm\ and \nustar.
This led to four flare peaks (marked with numbers 4, 5, 6 and 7).
We extracted strictly simultaneous spectra (0.4-78 keV) and performed a new broad-band spectroscopy.
We adopted the same models used to fit the time-averaged 0.4-78 spectrum  (Table~\ref{tab:simflares}).
When required by the presence of negative residuals, we report also the fit results
obtained with the inclusion of a  {\sc cyclabs} model,
although  in no case is the line significant.
We note that similar results were obtained using a Gaussian line in absorption ({\sc gabs}  in  {\sc xspec}) instead of {\sc cyclabs}, 
and a Comptonization continuum model ({\sc compTT}) instead of a PLCUT model.

\subsubsection{Not only flares: the lowest intensity state}

\begin{figure}
\begin{center}
\centerline{\includegraphics[width=6.0cm,angle=-90]{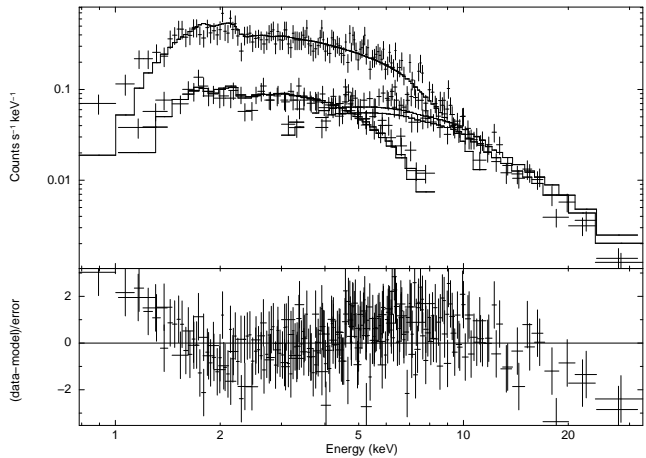}}
\vspace{0.0cm}
\caption{
Broad-band spectrum of the lowest intensity state. Counts spectra are shown together with the residuals
(in units of standard deviation) with respect to a simple 
absorbed power law, to clearly show the structures in the residuals.}
\label{fig:lowint_spec}
\end{center}
\end{figure}

To investigate the broad-band spectrum taken from the 
lowest intensity state where simultaneous \xmm\ and \nustar\ were available,
we accumulated a spectrum from the valleys just before and after flare n.~4. 
A meaningful spectroscopy could be performed in the energy band 0.8-30 keV.
Also in this low luminosity state (L$_X$$\sim$3$\times$10$^{35}$~erg~s$^{-1}$, 0.1-100 keV, see below), 
a fit with a single absorbed power-law revealed structured
residuals, with a low energy excess and a roll-over at high energy (Fig.~\ref{fig:lowint_spec}).
A simple double-component model, with a power law plus a blackbody (equally absorbed), already
provided a good fit ($\chi^{2}_{\nu}$/dof=0.939/267) to the emission (0.8-30 keV), 
resulting in an absorbing column density  N$_{\rm H}$=2$\times10^{22}$~cm$^{-2}$, $\Gamma$=1.3,
kT$_{\rm BB}$=1.3~keV and a blackbody radius R$_{\rm BB}$$\sim$150~m.
In order to compare the spectral properties of the low state with other intensity states,
we adopted both Model~1 and Model~2, fixing both Fe line properties and the power law cutoff to the time-averaged spectral shapes,
since in the low intensity state these latter were unconstrained.
The results are reported in Table~\ref{tab:low_spec}, to be compared with Table~\ref{tab:simflares} (flares) and Table~\ref{tab:av_spec} (time-averaged spectrum).

\begin{table}
        \centering
\caption[]{Broad-band spectrum during the lowest intensity state (0.8--30 keV).
}
\begin{tabular}{lcc}
 \hline
\noalign {\smallskip}
Parameter                                  &     Model~1                              &       Model~2       \\
\hline
\noalign {\smallskip}
N$_{\rm H}$      ($10^{22}$~cm$^{-2}$)     &    $1.52 ^{+0.37} _{-0.33}$                     &     $0.84 ^{+0.93} _{-0.77}$        \\
N$_{\rm H pcfabs}$ ($10^{22}$~cm$^{-2}$)  &       none                 &     $2.1 ^{+2.2} _{-0.7}$          \\
f$_{\rm  pcfabs}$                         &        none                &     $0.82 ^{+0.16} _{-0.52} $      \\
kT$_{\rm BB}$ (keV)                        &  $1.43 ^{+0.19} _{-0.15}$                &       none     \\
${\rm R_{BB}}$ (m)                         &    $280\pm{70}$                          &       none     \\
$\Gamma$$^a$                                  &  $0.78 ^{+0.18} _{-0.21}$              &       $0.98 ^{+0.09} _{-0.08}$              \\
E$_{cut}$ (keV)                              &  $9.7$ fixed                             &       $5.5 $ fixed                  \\
E$_{fold}$ (keV)                              &  $14.3$ fixed                           &        $15.3$ fixed     \\
E$_{line}$  (keV)                          &  $6.44$ fixed                                &        $6.45$ fixed          \\
$\sigma$$_{line}$ (keV)                    &  $0$ fixed                                   &   $0$ fixed            \\
Flux$_{line}$ (ph~cm$^{-2}$~s$^{-1}$)      &  3.2$\times10^{-5}$  fixed                  &      3.2$\times10^{-5}$  fixed      \\
EW$_{line}$$^b$  (eV)                      &  $<120$                                     &      $<120$     \\
Flux$^c$      (erg~cm$^{-2}$~s$^{-1}$)         &    (5.8$\pm{0.6}$) $\times$10$^{-11}$      &    (6.1$\pm{0.2}$) $\times$10$^{-11}$        \\
Luminosity$^c$  (10$^{35}$ erg~s$^{-1}$)       &    3.4$\pm{0.4}$                        &       3.6$\pm{0.1}$       \\
$\chi^{2}_{\nu}$/dof                           &    0.911/267                               &        0.911/267     \\
\hline
\label{tab:low_spec}
\end{tabular}
\tablecomments{
$^a$ Power-law photon index. \\
$^b$ Equivalent width. \\ 
$^c$ Fluxes (corrected for the absorption) and luminosities are in the energy range 0.1--100~keV.  The flux reported for  Model~2 is corrected for both 
absorbing components}
\end{table}

\subsubsection{Spin-phase resolved spectroscopy}
\label{sec:spinphasesel}

A spin-phase resolved spectroscopy  was performed on  strictly simultaneous \xmm\ and \nustar\ data.
From the hardness ratios reported in Fig.~\ref{fig:spinprofiles} a spin-phase resolved spectral extraction driven by the source hardness would suggest different spectral extractions, depending on the
energy bands adopted. So, we  simply divided the pulse period into five equally spaced intervals, and extracted five broad-band (0.4-78 keV) spectra. 
Note that we tried also narrower or broader spin-phase intervals, but we found that five is the best compromise to map eventual 
spectral variability, given the available counts statistics.

\begin{figure*}
\begin{center}
\begin{tabular}{cc}
\includegraphics[width=8.5cm,angle=0]{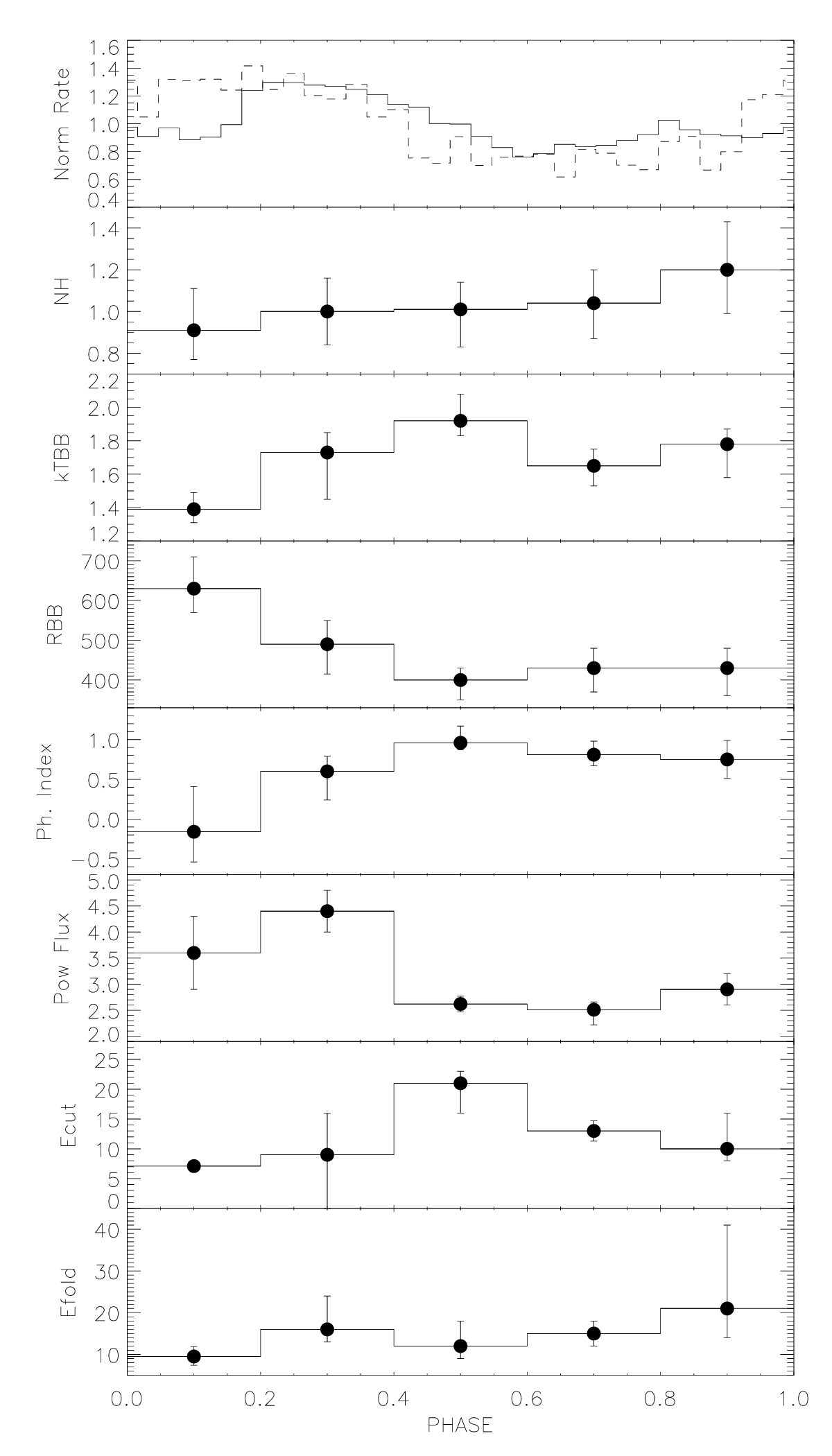} 
\includegraphics[width=8.5cm,angle=0]{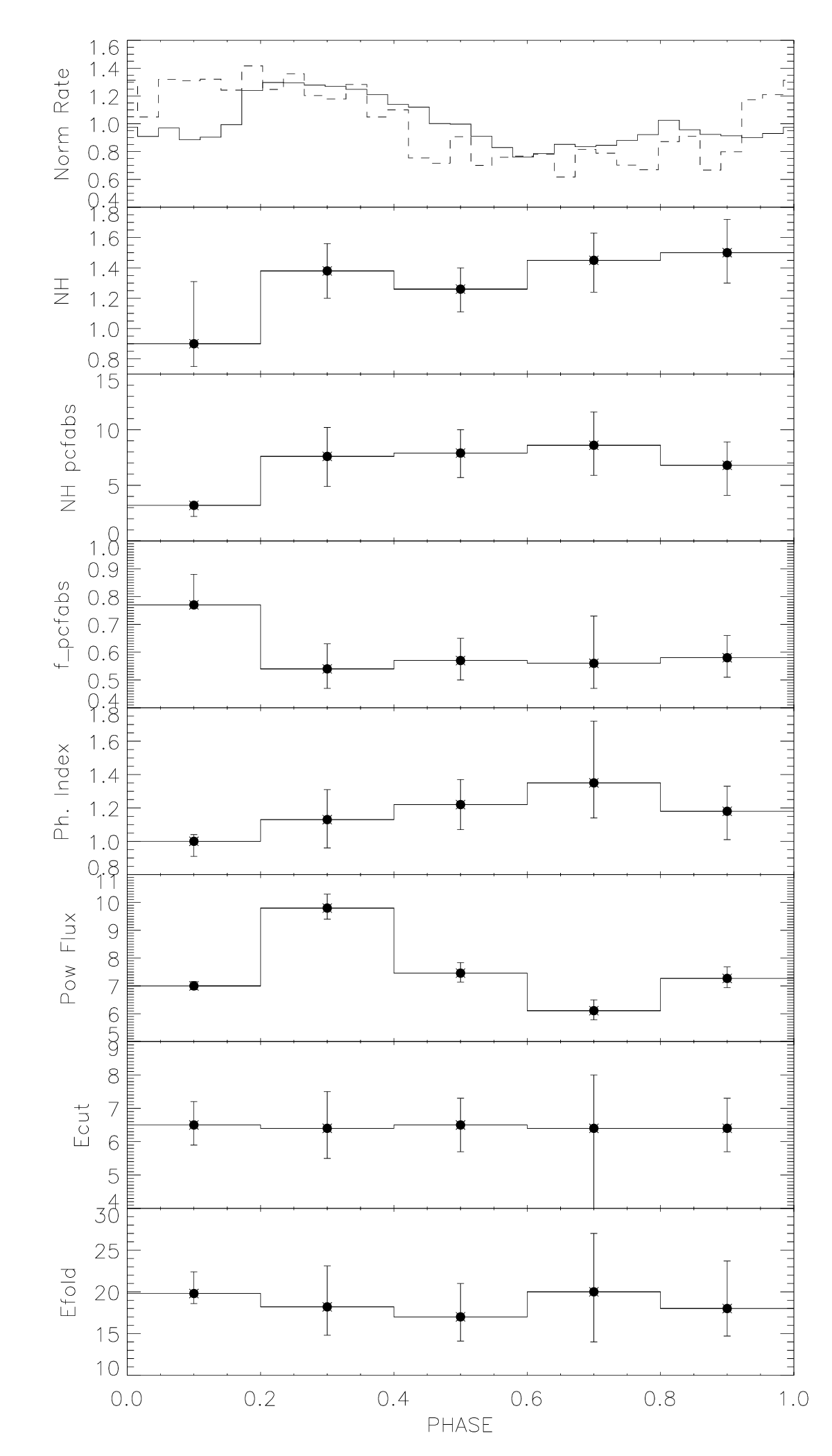}  \\
\end{tabular}
\end{center}
\caption{Results of the spin-phase-resolved spectroscopy of \xmm\ and \nustar\ simultaneous observations 
(the spectral parameters and their physical units are the same reported in Table~\ref{tab:spinphasespec}). 
{\em Left panel} shows the spectral paramenters with Model~1, while 
{\em Right panel} the results with Model~2. 
The top panels display the pulse profiles in two energy ranges for  comparison: 
0.4-12 keV (solid line) and 12-78 keV (dashed line). 
}
\label{fig:spinphasespec}
\end{figure*}

The same  models adopted before (Model~1 and Model~2) were applied to the five spectra  resulting in the
spectral parameters reported in Table~\ref{tab:spinphasespec} and summarized in Fig.~\ref{fig:spinphasespec}.
The evolution of the intensity along the pulse profile is dominated by the variability of the power-law flux (2-10 keV),
 while the spectral variability seems to be mainly due to the power-law slope, which
is the hardest in the first spin-phase interval ($\Delta\phi$=0.0-0.2), although the significant
variability of the power law cutoff energy and of the blackbody parameters (in Model 1) along the spin phase did not
allow us a straightforward interpretation.
 
\begin{figure}
\begin{center}
\begin{tabular}{cc}
\includegraphics[width=6.3cm,angle=-90]{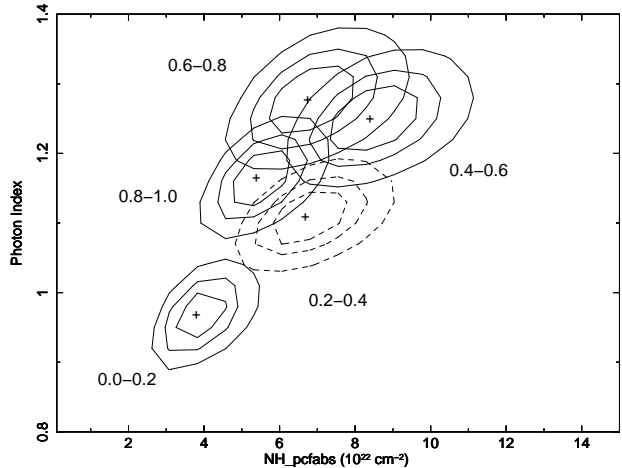} 
\end{tabular}
\end{center}
\caption{
Confidence contour levels (68\%, 90\% and 99\%) for  
the power law photon index and the additional absorbing column density partially
covering the X-ray emission along the spin phase. The resulting covering fractions in the  five spectra are consistent with each other (within 90\% uncertainty),
and their values are in the range 0.52-0.67. Numbers mark the five spin-phase-intervals. 
See Sect.~\ref{sec:spinphasesel}  for details.}
\label{fig:pcfabs_cont}
\end{figure}

When adopting Model~2, E$_{cut}$ and E$_{fold}$ parameters 
remained constant along the pulse profile. 
For this reason, we performed a second fit, fixing them to their mean value.  
We show in Fig.~\ref{fig:pcfabs_cont} the resulting confidence contour levels of the 
photon index and the additional absorbing column density partially
covering the power law. The numbers indicate the five spin-phase-intervals. 
In these fits, the covering fraction in the  five spectra was consistent with each other (within 90\% uncertainty), 
in the range 0.52-0.67. Again, spin-phase interval $\phi$=0.0-0.2 showed a harder power law emission. 

No significant negative residuals around 16-18~keV were found  during the spin-phase resolved spectroscopy.

\begin{table*}
        \centering
\caption[]{Broad-band (0.4-78 keV) spin-phase-resolved spectroscopy$^a$
}
\begin{tabular}{lccccc}

 \hline
Model~1 Parameters                                  &  $\Delta \phi$=0.0-0.2                  &  0.2-0.4                        &  0.4-0.6                           &  0.6-0.8                    &         0.8-1.0      \\
\hline\noalign {\smallskip}
N$_{\rm H}$      ($10^{22}$~cm$^{-2}$)     &  $0.91 ^{+0.20} _{-0.14}$               &  $1.0 \pm{0.16}$                &    $1.01 ^{+0.13} _{-0.18}$        &  $1.04 ^{+0.16} _{-0.17}$   &      $1.20 ^{+0.23} _{-0.21}$   \\
kT$_{\rm BB}$ (keV)                        &  $1.39 ^{+0.08} _{-0.10}$               & $1.73 ^{+0.28} _{-0.12}$        &   $1.92 ^{+0.09} _{-0.16}$         &   $1.65 ^{+0.12} _{-0.10}$  &      $1.78 ^{+0.20} _{-0.09}$     \\
${\rm R_{BB}}$ (m)                         &    $630 ^{+60} _{-80} $                 &     $490 ^{+75} _{-60}$         &    $400 ^{+50} _{-30}$             &    $430 ^{+60} _{-50}$      &         $430 ^{+70} _{-50}$        \\
$\Gamma$                                   &  $-0.16 ^{+0.38} _{-0.57}$              &  $0.60 ^{+0.36} _{-0.19}$       & $0.96 ^{+0.09} _{-0.21}$           &  $0.85 ^{+0.14} _{-0.17}$   &        $0.75 ^{+0.24} _{-0.24}$   \\
E$_{cut}$ (keV)                              &  $7.1^{+0.8} _{-0.9}$                 &  $9 ^{+7} _{-9}$                &  $21 ^{+2} _{-5}$                  &  $13.0 ^{+1.7} _{-1.7}$      &       $10 ^{+6} _{-2}$       \\
E$_{fold}$ (keV)                              &  $9.5 ^{+2.4} _{-2.1}$               &  $16 ^{+8} _{-3}$               &  $12 ^{+6} _{-3}$                  &  $15 ^{+3} _{-3}$            &       $21 ^{+20} _{-7}$        \\
Pow Flux$^b$ ($10^{-11}$~erg~cm$^{-2}$~s$^{-1}$)&    $3.6\pm{0.7}$         &    $4.4\pm{0.4} $               &       $2.62\pm{0.15} $             &  $2.51 ^{+0.15} _{-0.29}$   &       $2.9 ^{+0.3} _{-0.3}$     \\
$\chi^{2}_{\nu}$/dof                       &     0.988/782                            &  1.014/943                      &    1.011/749                       &  1.043/655                 &   0.997/752             \\
 \hline
Model~2 Parameters               &  $\Delta \phi$=0.0-0.2                 &  0.2-0.4                        &  0.4-0.6                           &  0.6-0.8                    &         0.8-1.0      \\
\hline\noalign {\smallskip}
 N$_{\rm H}$      ($10^{22}$~cm$^{-2}$)      &  $0.9 ^{+0.41} _{-0.15}$               &  $1.38 \pm{0.18}$                &    $1.26 ^{+0.14} _{-0.15}$        &  $1.45 ^{+0.18} _{-0.21}$   &      $1.50 ^{+0.22} _{-0.20}$   \\
 N$_{\rm H pcfabs}$ ($10^{22}$~cm$^{-2}$)   &  $3.2 ^{+1.0} _{-0.19}$                & $7.6 ^{+2.7} _{-2.6}$            &   $7.9 ^{+2.2} _{-2.1}$            &  $8.6 ^{+2.7} _{-3.0}$      &      $6.8 ^{+2.7} _{-2.1}$     \\
 f$_{\rm  pcfabs}$                          &    $0.77 ^{+0.01} _{-0.11} $           & $0.54 ^{+0.07} _{-0.09}$         &   $0.57 ^{+0.07} _{-0.08}$         &  $0.56 ^{+0.09} _{-0.17}$   &      $0.58 ^{+0.07} _{-0.08}$        \\
$\Gamma$                                        &  $1.00 ^{+0.09} _{-0.04}$          &  $1.13 ^{+0.17} _{-0.18}$       & $1.22 ^{+0.15} _{-0.15}$           &  $1.35 ^{+0.21} _{-0.37}$    &      $1.18 ^{+0.17} _{-0.15}$   \\
E$_{cut}$ (keV)                                 &  $6.5^{+0.7} _{-0.6}$              &  $6.4 ^{+1.1} _{-0.9}$          &  $6.5 ^{+0.8} _{-0.8}$             &  $6.4 ^{+1.6} _{-6.4}$      &       $6.4 ^{+0.89} _{-0.72}$       \\
E$_{fold}$ (keV)                                &  $19.8 ^{+2.6} _{-1.2}$             &  $18 ^{+5} _{-3}$              &  $17 ^{+4} _{-3}$                  &  $20 ^{+7} _{-6}$            &       $18 ^{+6} _{-3}$        \\
Pow Flux$^b$ ($10^{-11}$~erg~cm$^{-2}$~s$^{-1}$)&    $7.0 ^{+0.16} _{-0.13}$         &  $9.8 ^{+0.5} _{-0.4}$          &    $7.5 ^{+0.38} _{-0.32}$         &  $6.1 ^{+0.39} _{-0.33}$   &       $7.3 ^{+0.4} _{-0.3}$     \\
$\chi^{2}_{\nu}$/dof                       &    0.984/782                            &  1.027/943                      &    1.013/749                       &  1.046/655                 &   0.987/752             \\
\hline
\label{tab:spinphasespec}
\end{tabular}
\tablecomments{
$^a$ \xmm\ and \nustar\ strictly simultaneous spectra.  Model~1 is a double-component continuum  made of a blackbody plus a PLCUT model. 
Model~2 is a PLCUT model modified by a partial covering fraction absorption {\sc pcfabs} \\
$^b$Power-law flux (background corrected) in the energy band 2-10 keV. 
}
\end{table*}

\subsubsection{Hardness-ratio-selected spectroscopy}
\label{sec:hrsel}

\begin{figure}
\begin{center}
\centerline{\includegraphics[width=5.8cm,angle=-90]{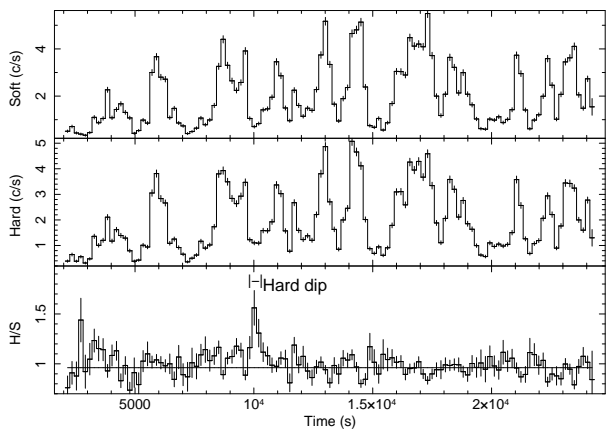}} 
\caption{\xmm\ EPIC pn light curves in two energy ranges (Soft = 1-4 keV; Hard = 4-12 keV), 
together with their hardness ratio (H/S) in the third panel. Bin time is 187~s. 
The only short time interval where a hardness
ratio variability is evident, has been marked in the bottom panel as ``hard dip'' and 
considered for further spectral analysis (Sect.~\ref{sec:hrsel}). 
The horizontal line is the fit to the hardness ratio with a constant model.}
\label{fig:hrsel}
\end{center}
\end{figure}

We investigated the spectral variability along the observation plotting different  hardness ratios versus time.
Whatever choice of hardness ratio (in both \xmm\ and \nustar\ data), 
some level of scatter around a mean value was always present, but with no clear 
trend suggesting a specific and meaningful hardness-ratio-selected spectroscopy.
The only exception was in case of the hardness ratio of 4-12 keV to 1-4 keV count rates (Fig.~\ref{fig:hrsel}): 
a hard peak lasting $\sim$500~s in the hardness ratio was found after 10$^{4}$~s from the start of the \xmm\ observation. 
We extracted an \xmm\ spectrum from this time interval (``hard dip'', hereafter), as it was not simultaneously covered by \nustar.
Meaningful spectroscopy was possible in the energy range 1-10 keV. 
A single absorbed power law model resulted in 
an absorbing column density  N$_{\rm H}$=(3.2$\pm{0.6}$)$\times10^{22}$~cm$^{-2}$, $\Gamma$=0.8$^{+0.16} _{-0.15}$,
and a flux corrected for the absorption F=3.9$\times10^{-11}$~erg~cm$^{-2}$~s$^{-1}$ (1--10 keV).
Adopting more complicated models (like Model~1 and Model~2) is meaningless, since several spectral parameters are unconstrained.

	      \section{Discussion}\label{sec:discussion}

We have reported on the joint observation of the SFXT \src\ with \xmm\ and \nustar, 
performed during the expected times of the 2016 February outburst.
These observations caught the source in outburst at its peak 
(L$_X$$\sim$10$^{36}$~erg~s$^{-1}$), as 
confirmed by the monitoring of the  light curve performed with \sw/XRT (Fig.~1).

The X--ray light curve  showed several bright flares with a dynamic range of one order of magnitude during the outburst.
The X--ray pulsar did not show evidence for spin period derivatives  with respect to previous
observations performed a decade ago \citep{Swank2007, Sidoli2007}. 
The rotational modulation is detected  for the first time also at hard energies with \nustar, with an energy-dependent spin
profile, evolving from an asymmetric main peak dominating at softer energies, to a single more symmetric peak at hard energies above $\sim$12~keV.

The source spectrum could be investigated for the first time simultaneously from 0.4 to 78 keV. It showed a complex structure that could not be described by a single 
absorbed power law, with a soft excess together with a cutoff at high energies.
Different spectral models were investigated, obtaining  a good fit  with a phenomenological double-component continuum composed by a blackbody together 
with a PLCUT model (Model 1). This description is  typically adopted in the X--ray spectroscopy 
of highly magnetized accreting pulsars \citep{Coburn2002}, where the power law is due 
to Comptonization in the accretion column and the blackbody could have different origins (see below).

We also explored an alternative deconvolution of the spectrum (a single-component model composed 
by a PLCUT), where the structures in the residuals at softer energies were accounted for by a partially covering absorption model, instead of a blackbody. 
This second model (Model 2)  resulted into an equivalently good description 
of the spectrum and it is worth reporting, since it is also
adopted in the literature when investigating spectra from HMXB pulsars 
(e.g. \citealt{Malacaria2016}). 
A narrow emission line is present in the time-averaged spectrum at an energy of 6.44 keV 
(consistent with fluorescent emission from almost neutral iron) 
and a low equivalent width of 40 eV. During other spectral selections, the faint emission line is not clearly 
visible, although it is still consistent with being there with a similar flux.
Fe lines with similar properties are normally found in the X-ray spectra of HMXB pulsars and are interpreted as 
produced by fluorescent emission in the wind of the supergiant donor illuminated by the X--ray pulsar \citep{Gimenez2015}. 

In both models, the PLCUT continuum 
resulted in a flat slope (with a harder photon index in Model 1 than in Model 2), and in energies of the cutoff
that are in agreement with what found in other X--ray pulsars spectra \citep{Coburn2002}. 
The blackbody emission in Model 1 resulted in a hot temperature of 1.6 keV from a small area of 460 meters (at a distance of
7 kpc). Such soft excesses have been discovered in different types of accreting pulsars (see \citealt{LaPalombaraMereghetti06} for a review),
with two different properties: hot temperatures (1-2 keV) with small emission radii of a few hundred meters,
versus colder blackbodies (0.1-0.5 keV) with a much larger emitting radius (R$_{BB}$$\sim$100 km). 
The blackbody component in \src\ is consistent with what found in low luminosity accreting pulsars 
(L$_X$$\sim$10$^{34}$-10$^{36}$~erg~s$^{-1}$), where
the hot and small blackbody is thought to be emitted 
from  the polar cap of the accreting pulsar \citep{LaPalombara2012}.

Comparing the X-ray flare spectra,
we rely on the broad-band \xmm\ and \nustar\ simultaneous spectroscopy (Table~\ref{tab:simflares}). 
In this case, only Model 1 provided evidence of variability in the spectral parameters with flux, where
harder slopes are found during brighter flare peaks (the faintest flare n.7 shows a significantly steeper power law),
again in agreement with what usually found in HMXB pulsars (e.g., \citealt{Walter2015, Martinez-Nunez2017} for reviews).
On the other hand, a similar trend in not present when using Model 2, where all parameters are consistent with being constant 
within the 90\% uncertainties. 
When compared with the spectrum extracted from the low intensity region of the light curve, only Model 1 resulted in
softer spectral parameters when the source is fainter (as usually observed in HMXB pulsars and SFXTs).

The absorption column density resulting from both models is always
in excess of the total Galactic (8$\times$10$^{21}$~cm$^{-2}$, \citealt{DL90}), consistent with 
local absorbing material, explained by the supergiant outflowing wind.
In Model~2, the power law continuum is partially covered (around 50\%) 
by an additional local absorbing column density. 
The physical meaning of a partial covering absorption might be 
ascribed to the clumpiness of the circumsource material, where the partial absorbers could be due to
 blobs in the donor wind. 
A denser wind clump passing in front of the X-ray source could also explain 
the hard dip lasting about 500~s caught during the uninterrupted \xmm\ observation.

During the spectroscopy of single flare peaks, we 
found a hint of an absorption feature at 17 keV that, if described with a cyclotron absorption
model, resulted in a line significance of 2.63~$\sigma$ (including trials)
derived by means of Monte Carlo simulations.
Given the low significance, the presence of this line cannot be claimed and needs further confirmation.
However, we note that if real, it is variable, since it is 
present neither in the time-averaged spectrum nor in the spin-phase selected spectroscopy.
This might resemble the behavior in another SFXT, IGRJ17544-2619, where a temporary cyclotron line was found.
The  cyclotron line at 16.8 keV  discovered 
in \nustar\ data by \citet{Bhalerao2015}, was not  confirmed by \citet{Bozzo2016} 
during another \nustar\ observation at  a similar source flux.

\subsection{The SFXT scenario}\label{sec:sfxt}

After ten years since SFXTs discovery, a huge progress has been made in understanding  their behaviour.

At first, it was proposed that Bondi-Hoyle direct accretion from dense clumps in the supergiant wind
could explain the sporadic flaring activity in SFXTs \citep{zand2005}.
A comparison of the clump properties implied by the SFXT X--ray light curves
with what predicted by the theory of line-driven hot stellar winds indicate 
that direct Bondi-Hoyle accretion from clumpy winds
cannot reproduce the X-ray data (see \citealt{Martinez-Nunez2017} and references therein for
a comprehensive review).
One of the difficulties of the clumpy wind scenario is that the accreted mass derived from the SFXT
X--ray flares (10$^{19}$-10$^{22}$~g) 
is typically larger than what expected from a single clump in stellar winds ($\sim$10$^{18}$~g).
The \src\ flare peak luminosity  implies an accreted mass of $\sim$10$^{19}$~g onto the NS, confirming once more this discrepancy. 
Note that this mass should be considered
as a lower limit to the clump mass, since a wind blob, if it is
larger than the accretion radius, might be only partially accreted by the pulsar.

Currently, two physical mechanisms might explain the SFXT flares in a  187~s pulsar like \src: 
either a centrifugal barrier \citep{Grebenev2007, Bozzo2008}),
or a quasi-spherical settling accretion regime  \citep{Shakura2012, Shakura2014}.
In the first case,  rescaling the \citet{Grebenev2007} equations to \src\
(spin period of 187 s and orbital period of 165 days) we derived a 
magnetic field of 1.2$\times$10$^{13}$~G needed to have an operating centrifugal barrier
(although we note that the equations reported by \citet{Grebenev2007} assume a circular orbit). 
If we adopt the average outburst X-ray luminosity of 10$^{36}$~erg~s$^{-1}$ instead of the wind parameters in this scenario, 
we found an even higher estimate for the magnetic field of $\sim$5$\times$10$^{13}$~G needed 
for the centrifugal barrier to operate. Note that also the source distance of 7 kpc is not accurate, 
but it should be considered as a lower limit \citep{Lorenzo2014}. 
In the second case, the settling accretion scenario sets in in slow pulsars 
(spin period $>$70 s) when the X--ray luminosity is below 4$\times$10$^{36}$~erg~s$^{-1}$ \citep{Shakura2012}.
Below this threshold the  matter captured inside the Bondi radius is not 
able to efficiently cool down by Compton processes, so that it cannot
penetrate the NS magnetosphere by means of Rayleigh-Taylor instabilities.
Hot matter accumulates above the NS magnetosphere forming a quasi-static shell
that cools down only by inefficient radiative cooling. 
\citet{Shakura2014} proposed that the bright SFXT flares are produced 
by the collapse onto the NS of this hot shell, triggered by reconnection events between 
the magnetic field
carried by the accreting matter and the neutron star magnetosphere.

Quasi-periodic flaring activity that has often been observed to punctuate the SFXT outbursts 
(e.g. SAX~J1818.6-1703, \citealt{Boon2016}), would be produced when
the shell, after a collapse, is replenished by new wind capture, 
so the flares repeat on a regular timescale, 
as long as the mass accretion into the magnetosphere is allowed.
The energy normally released in an SFXT bright flare ($\sim$10$^{39}$~erg) is in agreement
with estimated mass of the hot shell  \citep{Shakura2014}, that is also consistent with
the accreted mass we have derived here from the luminosity of the flares in \src\ ($\sim$10$^{19}$~g)

Our \xmm\ and \nustar\ observations of \src\ allowed us also to observe, for the first time in 
this source, X-ray flares repeating every $\sim$2-2.5 ks, a property that in the settling accretion scenario
is explained in a natural way. Note that the \src\ X--ray luminosity during outburst is fully compatible with
 the luminosity threshold predicted by \citet{Shakura2012} for the onset of the settling accretion. 
All these observational facts strongly favour an interpretation of the \src\ SFXT properties 
within the quasi-spherical settling accretion model.

Finally, we note that if the hint of a cyclotron line at 17 keV will be confirmed by
future observations (implying a NS surface magnetic field of  2$\times$10$^{12}$~G), 
it will reinforce the interpretation of the SFXT behavior within the settling accretion scenario.

\section{Conclusions}

We have performed for the first time a temporal and spectral analysis 
of the SFXT pulsar \src\  simultaneously in the broad-band 0.4-78 keV. 
Indeed, the broadest X--ray energy band previously available for this source 
was the 5--50~keV range, thanks to \inte\ observations \citep{SidoliPM2006}.
The spin period (187.0~$\pm{0.4}$~s) did not show evidence for variability with respect to previous observations. 
Pulsations were detected for the first time above 12 keV, showing an energy dependent spin profile. 
The broad-band spectra was successfully deconvolved adopting a double-component model made of a power law with a high energy cutoff  
together with a hot blackbody, which we interpret as emission from the polar caps of the NS. 
Alternatively, a partial covering model replacing the blackbody component resulted into an equally good description of the data.
The source light curve showed several bright flares spanning one order of magnitude in intensity, reaching 2$\times10^{36}$~erg~s$^{-1}$ 
at peak. The X--ray luminosity in outburst, together with the quasi-periodicity in the flaring activity
caught during the uninterrupted \xmm\ observation,  favour
a quasi-spherical settling accretion model as the physical mechanism explaining the SFXT behavior, 
although alternative possibilities (e.g. centrifugal barrier) cannot be ruled out.

\acknowledgements
This work is based on data from observations with \xmm, \nustar\ and \sw.
\xmm\ is an ESA science mission with instruments and
contributions directly funded by ESA Member States and the USA (NASA).
The \nustar\ mission is a 
project led by the California Institute of Technology,
managed by the Jet Propulsion Laboratory, and funded
by the National Aeronautics and Space Administration. 
This research made use of the {\em NuSTAR DAS} software
package, jointly developed by the ASDC (Italy) and Caltech (USA). 
We thank N. Gehrels and the Swift team for making the \sw\ ToO possible.
This work made use of data supplied by the UK \sw\ Science Data Centre at the University of Leicester. 
We thank A. De~Luca, S.~Molendi (for helpful suggestions during EPIC data reduction), 
A.~Bodaghee, F.~F\"urst, F.~Gastaldello, K.~Pottschmidt (for their useful advice on \nustar\ data), and
K.~Arnaud (for help with {\sc simftest}  in {\sc xspec}).
We are grateful to P.~Esposito, S.~Mereghetti and A.J.~Bird for interesting and helpful discussions.
We thank our anonymous referee for very constructive comments that helped to significantly improve the paper.
We acknowledge financial contribution from the grant from PRIN-INAF 2014 
``Towards a unified picture of accretion in High Mass X-Ray Binaries'' (PI Sidoli)
and from the agreement ASI-INAF NuSTAR I/037/12/0.



\bibliographystyle{aasjournal}

\end{document}